\documentclass[preprint,superscriptaddress,eqsecnum,nofootinbib]{revtex4}
\pdfoutput=1
\usepackage[T1]{fontenc} 
\usepackage{amssymb,amsbsy,bm,amsmath,psfrag,dsfont}
\usepackage{hyperref}
\usepackage{graphicx}
\usepackage{epsfig}
\usepackage{subfigure}
\usepackage{graphicx,amssymb,amsbsy,bm,amsmath,psfrag,dsfont,rotating}

\newcommand{\beq}[1]{\begin{equation}\label{#1}}
\newcommand{\eeq}{\end{equation}}
\newcommand{\ba}{\begin{array}}
\newcommand{\ea}{\end{array}}

\newcommand{\be}{\begin{equation}}
\newcommand{\ee}{\end{equation}}
\newcommand{\bea}{\begin{eqnarray}}
\newcommand{\eea}{\end{eqnarray}}

\newcommand{\al}{\alpha}

\newcommand{\numu}{\nu_{\mu}}
\newcommand{\nutau}{\nu_{\tau}}

\begin{document}

\title{\boldmath Flavor Ratios and Mass Hierarchy at Neutrino Telescopes}

\author{Lingjun Fu}\email{lingjun.fu@vanderbilt.edu}
\affiliation{Department of Physics and Astronomy, Vanderbilt University, Nashville, TN 37235, USA}
\author{Chiu Man Ho}\email{cmho@msu.edu}
\affiliation{Department of Physics and Astronomy, Michigan State University, East Lansing, MI 48824, USA}

\date{\today}
\begin{abstract}
The observation of high-energy extraterrestrial neutrinos at IceCube represents the beginning of the era of neutrino astronomy. In this paper,
we study the cosmic neutrino flavor ratios against the Dirac CP-violating phase at neutrino telescopes,
taking into account the charged-current and neutral-current interactions at the detectors. We then demonstrate how to probe mass
hierarchy at future neutrino telescopes by the precise measurements of the cosmic neutrino flavor ratios. We show that the sensitivity of our
scheme is independent of the undetermined values of the Dirac CP-violating phase. We also explore the possible effects of
active-sterile mixing, neutrino decay and pseudo-Dirac nature of neutrinos.
\end{abstract}

\maketitle

\section{Introduction}

The terrestrial neutrino experiments have been making significant progress towards determining the neutrino properties.
For instance, the magnitude of the mass-squared splittings and the mixing angles $\theta_{12}, \theta_{23}$ have been relatively well measured.
For years, the neutrino mixing data have been consistent with $\theta_{13}=0$. This accommodates the $\numu$-$\nutau$ symmetry naturally realized by
the TriBimaximal (TBM) model~\cite{TBM}. However, DAYA-BAY \cite{DAYA-BAY} and RENO \cite{RENO} have recently observed
$\sin^2\,(2 \theta_{13}) = 0.092 \pm 0.016\, (\textrm{stat.}) \pm 0.005\, (\textrm{syst.})$ and
$\sin^2\,(2 \theta_{13}) = 0.113 \pm 0.013\, (\textrm{stat.}) \pm 0.019 \,(\textrm{syst.})$ at 68\% C.L. respectively.
This disfavors the TBM model and represents yet another important step towards the complete understanding of the neutrino sector.

Despite the significant progress made by the experiments, neutrinos remain mysterious. We are still ignorant of some basic
neutrino properties: Is the neutrino mass hierarchy normal or inverted? Are neutrinos Dirac or Majorana in nature? What is the absolute mass
scale of neutrinos? What is the Dirac CP-violating phase? Each of these questions is important on its own. The focus of this paper is
the neutrino mass hierarchy.

Currently, there are a few relatively promising experiments proposed to measure the neutrino mass hierarchy.
These include LBNE (accelerator) \cite{LBNE}, PINGU (atmospheric) \cite{PINGU} and JUNO (reactor) \cite{JUNO}.
The timescale of these experiments ranges from 2025 to 2030 for the first results \cite{WhitePaper}. The sensitivities
of these experiments are quantified in \cite{Huber,Ge:2013zua,Ge:2013ffa}.

Recently, the IceCube collaboration has reported an excess of 37 events relative to the atmospheric neutrino
background \cite{IceCube2,IceCube26,IceCube37}. Apart from the two events that are almost certainly produced in cosmic-ray air
showers, 3 events (among the remaining 35) have energies slightly above PeV while the other
32 events have energies between 20 TeV and 400 TeV. The overall signal significance is that the analysis rejects a purely
atmospheric explanation of these events at 5.7$\sigma$.
The hope is that after an ensemble of neutrino events have been collected, track-topologies will
allow one to reveal the neutrino flavor ratios arriving on Earth~\cite{MeasuringFlavor}.

In this paper, we study the cosmic neutrino flavor ratios against the Dirac CP-violating phase at neutrino telescopes,
taking into account the charged-current and neutral-current neutrino-nucleon interactions at the detectors.
Then, we propose that precise measurements of the cosmic neutrino flavor ratios at neutrino telescopes may provide yet another possible
way of determining the neutrino mass hierarchy. As we shall see, the sensitivity of our scheme is independent of the undetermined
values of the Dirac CP-violating phase.



\section{Cosmic Neutrino Flavor Ratios at Neutrino Telescopes}

In the standard treatment of neutrino oscillations, neutrino flavor states and mass eigenstates are
related by a unitary transformation: $|\nu_\al \rangle =\sum_{j}\, U_{\al j}^\ast\, |\nu_j \rangle $\,,
where $\al=e,\,\mu,\,\tau$ and $j=1,\,2,\,3$ are the indices for the flavor states and mass eigenstates respectively.
This unitary transformation is described by the Pontecorvo-Maki-Nakagawa-Sakata (PMNS) matrix $U$ with
the elements $U_{\al j}=\langle \nu_\al | \nu_j\rangle$:
\bea
\left(
      \begin{array}{ccc}
        U_{e 1} & U_{e 2} & U_{e 3} \\
        U_{\mu 1} & U_{\mu 2} & U_{\mu 3} \\
        U_{\tau 1} & U_{\tau 2} & U_{\tau 3} \\
      \end{array}
    \right) =\left(
               \begin{array}{ccc}
                 c_{21}\,c_{13} & s_{21}\,c_{13} & s_{13}\,e^{-i\,\delta} \\
                 -s_{21}\,c_{32} -c_{21}\,s_{32}\,s_{13}\,e^{i\,\delta} & c_{21}\,c_{32}-s_{21}\,s_{32}\,s_{13}\,e^{i\,\delta} & s_{32}\,c_{13} \\
                 s_{21}\,s_{32} -c_{21}\,c_{32}\,s_{13}\,e^{i\,\delta} & -c_{21}\,s_{32} -s_{21}\,c_{32}\,s_{13}\,e^{i\,\delta} & c_{32}\,c_{13} \\
               \end{array}
             \right)
\eea
where $c_{jk}=\cos(\theta_{jk})$, $s_{jk}=\sin(\theta_{jk})$ and $\delta$ is the Dirac CP-violating phase.
After propagating over distance $L$, the flavor state $|\nu_\al \rangle $ evolves into
$|\nu_\al (L) \rangle =\sum_{k}\, e^{-i\,E_k\,L}\,U_{\al k}^\ast\, |\nu_k \rangle$.
The transition probability of $|\nu_\al (L)\rangle \rightarrow |\nu_\beta \rangle$ is then given by
$P_{\al\beta}= |\langle \nu_\beta|\nu_\al(L) \rangle|^2$ for any $\al,\,\beta=e,\,\mu,\,\tau$.

For cosmic neutrinos, the characteristic propagation distance is much larger than the oscillation length. Thus, we
can perform a statistical average over a neutrino ensemble. This eliminates the quantum-mechanical phase
$\phi_{jk}\,\equiv \,L\,(m_j^2-m_k^2)/2E$ between states, leaving a simple propagation matrix $P$:
\bea
\label{Pab}
P_{\alpha\beta} &=& \sum_j \, | U_{\alpha j} |^2\, | U_{\beta j} |^2 = \left(\,|U|^2\, \left(\,|U|^2\,\right)^T\,\right)_{\alpha\beta} \\
 &=& \left(
      \begin{array}{ccc}
        P_{ee} & P_{e\mu} & P_{e\tau} \\
        P_{\mu e} & P_{\mu\mu} & P_{\mu\tau} \\
        P_{\tau e} & P_{\tau \mu} & P_{\tau \tau} \\
      \end{array}
    \right)\,,
\eea
where $|\,U\,|^2$ is given by
\bea
\label{Usquared}
 |U|^2 \equiv \left(
              \begin{array}{ccc}
                |U_{e 1}|^2 ~& ~|U_{e 2}|^2 ~&~ |U_{e 3}|^2 \\ \\
                |U_{\mu 1}|^2 ~&~ |U_{\mu 2}|^2 ~&~ |U_{\mu 3}|^2 \\ \\
                |U_{\tau 1}|^2 ~&~ |U_{\tau 2}|^2 ~&~ |U_{\tau 3}|^2 \\
              \end{array}
            \right)\,.
\eea
We can understand the propagation matrix $P$ as follows. The physically relevant basis for neutrino propagation is the mass basis in which the
mass eigenstates have definite masses. Since the mass eigenstates labeled by $j$ are unobserved, we need to sum over them. Besides,
each of these mass eigenstates should be weighted by its classical probability $|U_{\alpha j}|^2$ to overlap with $|\nu_\alpha \rangle$ produced
at the source, times its classical probability $|U_{\beta j}|^2$ to overlap with $|\nu_\beta \rangle$ detected on Earth.
Since phase-averaging eliminates the quantum-mechanical phase $\phi_{jk}$ and thereby restoring the CP-invariance, the matrix $P$ describes
both neutrino and anti-neutrino propagations equally. Furthermore, according to the CPT-theorem, CP-invariance implies T-invariance. This means
that the matrix $P$ is also symmetric, namely $P_{\alpha\beta} = P_{\beta\alpha}$.

For a given neutrino flavor ratio unit-vector ${\vec W} \equiv (W_e, W_\mu, W_\tau)$ produced at cosmic sources,
the corresponding flavor ratio ${\vec \phi}\equiv (\phi_e, \,\phi_\mu, \,\phi_\tau)$ measured on Earth can be obtained from
\beq{FlavorProp1}
{\vec \phi} = P\,{\vec W}\,.
\eeq
Due to non-zero $\theta_{13}$, the $\nu_\mu-\nu_\tau$ symmetry is broken and $P$ is now invertible. This means that measurements
of \,${\vec \phi}$\, on Earth can now be used to directly reveal ${\vec W}$ through the relation ${\vec W} = P^{-1}\,{\vec \phi}$\,
\cite{Fu,Xing:2006xd}.

Since $\tau$ are not strongly suppressed at cosmic sources, the initial neutrino flavor compositions generally do not have $\nu_\tau$ \cite{Lipari,PRWei,Choubey:2009jq,Barger:2014iua}.
Although $\nu_\tau$ may be produced from charmed meson decays, the production of charmed mesons requires a higher energy threshold and it has
a lower cross-section. This means that the amount of $\nu_\tau$ produced from this channel is negligible \cite{WaxmanBahcall,Xing:2006uk}. Therefore, it is
reasonable to parameterize the most general injection model as
\bea
\label{general}
\left(\,W_e\,:\,W_\mu\,:\, W_\tau\,\right) = \left(\,\alpha\,:\,1 -\alpha \,:\, 0\,\right)\,,
\eea
where $\alpha$ is a free parameter ranging from 0 to 1.

Neutrino telescopes are particularly adept at distinguishing the muon tracks from the showering events.
Thus, an experimentally useful observable would be the track-to-shower ratio:
\bea
\label{DefineR}
R = \frac{p_{{\rm CC}}\;\phi_\mu}{p_{{\rm NC}}\;\phi_\mu + \phi_e+\phi_\tau}\,,
\eea
where $p_{{\rm CC}}$ and $p_{{\rm NC}}$ are the probabilities of charged-current (CC) and neutral-current (NC) neutrino-nucleon interactions
respectively. For both $\nu_\mu$ and $\bar{\nu}_\mu$, the charged-current processes contribute to the track events while the neutral-current processes contribute to the shower events. There will be background events contributing to each of the track and shower events. Hence, $R$ represents
the track-to-shower ratio to be observed by neutrino telescopes with the background events subtracted. In general, $p_{{\rm CC}}$ and $p_{{\rm NC}}$
are energy-dependent. Around 100 TeV, $p_{{\rm CC}}$ and $p_{{\rm NC}}$ stay relatively constant and
we have $p_{{\rm CC}} \approx 0.72$ and $p_{{\rm NC}} \approx 0.28$ for both $\nu_\mu$ and $\bar{\nu}_\mu$ (see Tables I and II in \cite{Gandhi}).
We will take these values for the rest of our study. Notice that for $\nu_\tau$ and $\bar{\nu}_\tau$ with energies above a few PeV, about $20\%$ of their CC interactions will also contribute to the track events through the ``double-bang" events \cite{DoubeBang}, and Eq. \eqref{DefineR} will need to be modified correspondingly. However, we have explored this modification and found that it does not change the qualitative results in the current paper. Also, most of the neutrino events observed by IceCube so far have energies below PeV, and so we will just present our results using Eq. \eqref{DefineR}.

For the most general case in Eq. \eqref{general}, we obtain (using $p_{{\rm CC}}+p_{{\rm NC}}=1$)
\bea
\label{generalR}
R = \frac{p_{{\rm CC}}\,\left[\,P_{\mu e}\,\alpha + P_{\mu \mu}\, \left(\,1-\alpha\,\right)\,\right]}
{\left[\,1- p_{{\rm CC}}\,P_{\mu e} \,\right]\,\alpha
+ \left[\,1- p_{{\rm CC}}\,P_{\mu \mu} \,\right] \,\left(\,1-\alpha\,\right)}\,.
\eea
Currently, there are three popular models for the production of cosmic
neutrinos.\footnote{A thorough overview of neutrino injection models is provided by \cite{Hummer:2010ai}.} They are:
\begin{itemize}
  \item (1) Pion-Chain: ~
  Neutrinos could be created from hadronic sources such as
  $ p + p \rightarrow \pi^{+} \rightarrow \mu^+ + \nu_\mu \rightarrow e^+ + \nu_e + \nu_\mu + \bar{\nu}_\mu $ or
  $p + p  \rightarrow \pi^{-} \rightarrow \mu^- + \bar{\nu}_\mu \rightarrow e^- + \bar{\nu}_e + \nu_\mu + \bar{\nu}_\mu$.
  The high-energy $\pi^+$ could also be produced from the interactions between accelerated protons and photons.
  Both cases lead to $(W_e\,:\, W_\mu\,:\,W_\tau) = (\frac13\,:\,\frac23\,:\, 0)$. This is a special case of Eq. \eqref{generalR}
  with $\alpha =1/3$:
  \bea
  R = \frac{p_{{\rm CC}}\,\left(\,P_{\mu e} + 2\,P_{\mu \mu}\,\right)}
  {\left[\, 1- p_{{\rm CC}}\, P_{\mu e} \,\right]  + 2\,\left[\, 1- p_{{\rm CC}}\,P_{\mu \mu} \,\right]}\,.
  \eea
  As a remark, in the TBM model \cite{TBM}, we have
  \bea
  P_{\textrm{TBM}} = \frac{1}{18}\, \left(
                                      \begin{array}{ccc}
                                        10 & 4 & 4 \\
                                        4 & 7 & 7 \\
                                        4 & 7 & 7 \\
                                      \end{array}
                                    \right)\,,
  \eea
  which implies that $(\phi_e\,:\, \phi_\mu\,:\,\phi_\tau)_{\textrm{TBM}} = (\frac13\,:\,\frac13\,:\,\frac13)$.
  Since the $\nu_\mu-\nu_\tau$ symmetry is slightly broken, we expect the actual
  $(\phi_e\,:\, \phi_\mu\,:\,\phi_\tau)$ to deviate slightly from $(\frac13\,:\,\frac13\,:\,\frac13)$.
  \item (2) Damped-Muon: ~
  In the pion decay chain mentioned above, it is possible that the flux of muons gets depleted. This may happen if the muons lose energy in a
  strong magnetic field or get absorbed in matter \cite{MeszarosWaxman}. This leads to $(W_e\,:\, W_\mu\,:\,W_\tau) = (0\,:\,1\,:\, 0) $ which is
  a special case of Eq. \eqref{generalR} with $\alpha = 0$:
  \bea
  R = \frac{p_{{\rm CC}}\;P_{\mu \mu}}{1- p_{{\rm CC}} \,P_{\mu \mu} }\,.
  \eea
  \item (3) Beta-Beam: ~
  Some sources may dominantly emit neutrons. These neutrons could be produced from the photo-dissociation of heavy nuclei \cite{beta-beam}
  or the interactions between accelerated protons and photons \cite{Neutrons}.
  The decays of these neutrons $(n \rightarrow p + e^- + \bar{\nu}_e)$ lead to $(W_e\,:\, W_\mu\,:\,W_\tau) = (1\,:\,0\,:\, 0) $
  which is a special case of Eq. \eqref{generalR} with $\alpha = 1$:
  \bea
  R = \frac{p_{{\rm CC}}\;P_{\mu e}}{1- p_{{\rm CC}} \,P_{\mu e} }\,.
  \eea
\end{itemize}

\section{Flavor Ratios and Mass Hierarchy: ~ Standard Scenario}

In this section, we study the standard scenario with three active neutrinos. We first illustrate our idea with the three popular
injection models, and then consider the most general injection model parameterized by Eq. \eqref{general}.

Throughout the entire discussion, we embrace the most updated global best-fit data of three neutrino mixing \cite{Valle:2014} for
normal hierarchy (NH) and inverted hierarchy (IH):
\bea
\sin^2 \theta_{13} &=& 0.0234, ~~ \sin^2 \theta_{32}=0.567/0.467, ~~ \sin^2 \theta_{12}=0.323 ~~~~~~~  (\,\textrm{NH}\,)\,, \\
\sin^2 \theta_{13} &=& 0.0240, ~~ \sin^2 \theta_{32}=0.573, ~~ \sin^2 \theta_{12}=0.323 ~~~~~~~ (\,\textrm{IH}\,)\,.
\eea
Therefore, according to the best-fit analysis in \cite{Valle:2014}, the neutrino mass hierarchy is related to $\theta_{23}$.
This suggests three possible cases:
(1) normal hierarchy with $\sin^2 \theta_{32}=0.567$ (NH1), \,
(2) normal hierarchy with $\sin^2 \theta_{32}=0.467$ (NH2), \,
(3) inverted hierarchy with $\sin^2 \theta_{32}=0.573$ (IH).

Since neutrino telescopes are particularly adept at distinguishing the muon tracks from the showering
events, the main observable to be studied in this paper is $R$. The role of the uncertainties of the neutrino mixing parameters in flavor measurements at neutrino telescopes have been discussed before \cite{Meloni:2006gv,Esmaili:2009dz}. Now, DAYA-BAY \cite{DAYA-BAY} and
RENO \cite{RENO} have already provided us with the precise value for $\theta_{13}$. As far as neutrino oscillation is
concerned, the only unknown parameter in the PMNS matrix $U$ is the Dirac CP-violating phase $\delta$ \cite{Meloni:2012nk}. Thus,
we will plot $R$ against $\delta$ to see the dependence of the sensitivity on the-only-unknown parameter $\delta$.

\subsection{Three Popular Injection Models }

\begin{figure}[t!]
\includegraphics[height=6.5cm, width=7.5cm]{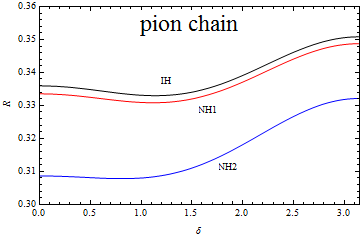}
~~
\includegraphics[height=6.5cm, width=7.5cm]{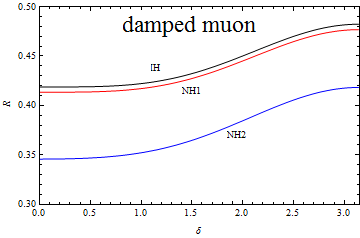}
~~~~~~
\begin{center}
\includegraphics[height=6.5cm, width=7.5cm]{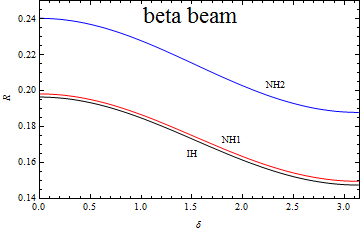}
\end{center}
\caption{~$R$ against $\delta$ for pion-chain, damped-muon and beta-beam injection models.}
\label{fig:SM}
\end{figure}

In each of the three popular injection models, we calculate $R$ for NH1, NH2 and IH. The three figures in
Fig. \ref{fig:SM} display $R$ against $\delta$ for pion-chain, damped-muon and beta-beam injection models respectively.
From Fig. \ref{fig:SM}, it is obvious that the fluctuation of $R$ with varying $\delta$
is small in the pion-chain case while relatively large in the other two cases. For instance, we have $ 0.3 < R < 0.35$ for pion-chain,
$ 0.35< R < 0.5$ for damped-muon and $ 0.14 < R < 0.25$ for beta-beam.
In particular, completely independent of $\delta$, NH1, NH2 and IH, these three injection models lead to distinctive ranges of $R$.
This interesting feature allows us to distinguish between these three injection
models in the near future when neutrino telescopes have observed statistically sufficient number of events such that a conclusive value of
$R$ could be established.

In all of the three injections models, it is difficult to distinguish NH1 from IH. However, the differences between NH2 and IH in these
injection models could be more significant. For instance, in the pion-chain case, the difference between NH2 and IH is at least 0.02.
The typical differences between NH2 and IH in damped-muon and beta-beam cases are 0.06 and 0.04 respectively. Most importantly, the magnitudes of
the differences between NH2 and IH in all of these three injection models are almost independent of the undetermined values of $\delta$.

Therefore, when the neutrino telescopes can achieve the sensitivities down to about 0.02 or lower, we may be able to probe the mass
hierarchies NH2 and IH by measuring the cosmic neutrino flavor ratios at the detectors. It is noteworthy that this scheme does not depend
on a precise measurement of the Dirac CP-violating phase. The prelude to probing mass hierarchy by cosmic neutrino flavor ratios is the
determination of the relevant injection model by establishing a conclusive value for $R$ at neutrino telescopes.

\subsection{The General Injection Model}

\begin{figure}[t!]
\centering
\includegraphics[height=7cm, width=8cm]{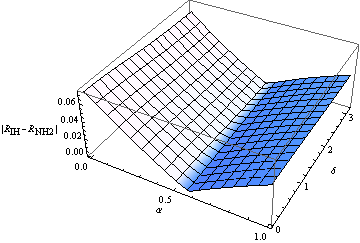}
\caption{The difference $|R_{\textrm{IH}} - R_{\textrm{NH2}}|$ as a function of both $\alpha$ and $\delta$.}
\label{fig:3Ddiff}
\end{figure}

In reality, it is possible that there are some deviations from pion-chain, damped-muon and beta-beam injection models which have
exact initial neutrino flavor compositions. Of course, if the deviations from these three popular injection models are
perturbatively small, then the previous results and conclusions would be sufficiently reliable. We don't know yet whether the deviations
are small, so it is useful to study the general injection model parameterized by Eq. \eqref{general} as well. With one more free
parameter $\alpha$ now, we will display a 3D plot with x-axis, y-axis and z-axis being $\alpha$, $\delta$ and $R$ respectively.

Since we are interested in the prospects of using the cosmic neutrino flavor ratios to probe mass hierarchy, it would be illuminating
to investigate the difference $|R_{\textrm{IH}} - R_{\textrm{NH2}}|$ as a function of both $\alpha$ and $\delta$. We neglect
the difference $|R_{\textrm{IH}} - R_{\textrm{NH1}}|$ because it is close to zero. In Fig. \ref{fig:3Ddiff},
we see that $|R_{\textrm{IH}} - R_{\textrm{NH2}}|$ can be as large as 0.07. The magnitude depends
mainly on $\alpha$  and is almost independent of $\delta$. It is especially small when the injection model has roughly
equal $\nu_e$ and $\nu_\mu$ initial compositions ($\alpha \approx 1/2$). Hence, unless $\alpha$ is close to 1/2, neutrino telescopes
will have the potential to distinguish NH2 from IH when they achieve the sensitivities down to about 0.02 or lower.

Currently, both T2K \cite{T2K} and NO$\nu$A \cite{NOVA} are trying to measure $\delta$. It is possible that a reliable value for
$\delta$ is ready by the time when neutrino telescopes have observed statistically sufficient number of events to establish a conclusive value for
$R$. If so, we could then reduce the 3D plots to 2D plots with $R$ against $\alpha$. (Actually, since $|R_{\textrm{IH}} - R_{\textrm{NH2}}|$ depends
mainly on $\alpha$ and is almost independent of $\delta$, we could have plotted $R$ against $\alpha$ with $\delta$ fixed to be a random value. While
this might be sufficiently illuminating, we kept those 3D plots for precise analyses.) Again, when neutrino telescopes have acquired sufficient
sensitivities, they will be able to probe the mass hierarchy for the general injection model.


\subsection{Caveats}

In the analyses conducted above, we have adopted the recent global best-fit data of three-neutrino mixing provided
by \cite{Valle:2014}. The data suggest three possible cases, namely NH1, NH2 and IH. Their $\pm 1 \sigma$ data also indicate
a preference for IH with $\theta_{23} > \pi/2$. The corresponding numbers for NH2 and IH
in \cite{Valle:2014} are consistent with those in \cite{Schwetz}. Based on this feature, we have shown that neutrino telescopes can distinguish
between NH2 and IH once they have reached the sufficient sensitivities.
Apparently, both of the $\pm 2 \sigma$ and $\pm 3 \sigma$ data in \cite{Valle:2014} do not share this feature. This means that our scheme for probing mass hierarchy at neutrino telescopes relies on the validity of the best-fit data.
However, if this feature persists in the forthcoming more precise global neutrino data-fittings, our scheme will remain a possible one.

\section{Flavor Ratios and Mass Hierarchy: ~ Beyond Standard Scenario}

\subsection{Active-Sterile Mixing}

Short baseline neutrino experiments such as LSND \cite{LSND} and MiniBooNE \cite{MiniBooNE} seem to suggest the
existence of eV-scale sterile neutrinos. Although the stringent bound from PLANCK satellite \cite{PLANCK}, taken at its face value,
disfavors eV-scale sterile neutrinos, there are promising ways to reconcile their existence with cosmology \cite{Reconcile1,Reconcile2}.
So it would be interesting  to study the active-sterile mixing scenario.

To include the eV-scale sterile neutrinos, we adopt the parameterization and fits for the minimal 3+2 neutrino
model found in \cite{Donini}. It is quite straightforward to extend the $3 \times 3$ case to $5 \times 5$ one with the new
fit values of $| U_{\alpha j} |$ plugged in:
\bea
\label{sterile1}
| U_{e 4}| &=& 0.149, ~~ | U_{e 5}|=0.127, ~~ | U_{\mu 4}|=0.112, ~~ | U_{\mu 5}|=0.127  ~~~~~~ (\,\textrm{NH}\,)\,, \\
\label{sterile2}
| U_{e 4}| &=& 0.139, ~~ | U_{e 5}|=0.122, ~~ | U_{\mu 4}|=0.138, ~~ | U_{\mu 5}|=0.107  ~~~~~~ (\,\textrm{IH}\,)\,.
\eea
However, the contributions from these new extra terms to the original values
in $P_{\alpha \beta}$ are only of order $\sin^4 \theta_{13}$. These are negligible compared to
the original $P$ matrix elements at the order of $\sin \theta_{13}$ \cite{Fu}. Therefore, our scheme is not affected by the
active-sterile mixing.

\subsection{Neutrino Decay}

\begin{figure}[t!]
\centering
\includegraphics[height=6.5cm, width=7.5cm]{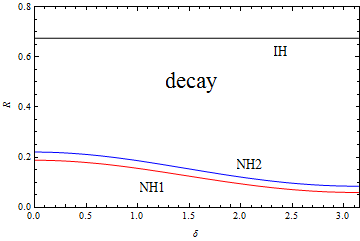}
\caption{~$R$ against $\delta$ for neutrino decay.}
\label{fig:decay}
\end{figure}

It is possible that neutrinos decay in the following manner \cite{nuDK,Baerwald:2012kc}:
\bea
\nu_i \, \rightarrow \, \nu_j + X ~~~~ \textrm{and} ~~~~ \nu_i \rightarrow \bar{\nu}_j  + X \,,
\eea
where $X$ is a very light or massless particle such as a Majoron. Viable Majoron models leading to neutrino decays have been discussed
in \cite{Majoron}. Neutrino decay has recently been invoked to explain the apparent deficit of $\nu_\mu$ events predicted by the pion-chain and
damped-muon injection models at IceCube \cite{Pakvasa,Valle,Mena}. However, this deficit was not statistically
significant, and a more careful theoretical analysis \cite{Chen:2013dza} was also suggested. Most recently, a better analysis by the
IceCube collaboration \cite{IceCubeBetter} indicates that this scenario is not supported by the data.

The value of $R$ will be greatly altered if cosmic neutrinos from distant astrophysical sources decay.
For simplicity, we assume that all the decays are complete and there is no other new physics besides decay. Regardless of any injection models,
the final remnants are $\nu_1$ in NH and $\nu_3$ in IH. Thus, one can easily get \cite{decay}:
\bea
R &=& \frac {p_{{\rm CC}}\,| U_{\mu 1}|^2} {p_{{\rm NC}}\,| U_{\mu 1}|^2 +| U_{e 1}|^2+| U_{\tau 1}|^2} ~~~~~~~~ (\,\textrm{NH}\,)\,,\\
R &=& \frac {p_{{\rm CC}}\,| U_{\mu 3}|^2} {p_{{\rm NC}}\,| U_{\mu 3}|^2 +| U_{e 3}|^2+| U_{\tau 3}|^2} ~~~~~~~~ (\,\textrm{IH}\,)\,.
\eea
From Fig. \ref{fig:decay}, it is clear that if we observe track-event dominated ratio (\,$R \sim 0.65$\,), it would strongly
indicate neutrino decay with IH, regardless of the undetermined values of $\delta$. If we observe
$0.25 < R < 0.65 $, neutrino decay is disfavored. An observation of a shower-event dominated ratio (\,$ 0.05 <R < 0.25$\,)
may favor neutrino decay with NH. Recall that the beta-beam injection model (see Fig. \ref{fig:SM}) predicts $0.14 < R < 0.25$ for
both of NH and IH. Thus, neutrino decay with NH would be strongly favored if we observe $0.05 < R < 0.14$.


\subsection{Pseudo-Dirac Neutrinos}

Neutrinos may be pseudo-Dirac states such that each generation is actually composed of
two maximally-mixed Majorana neutrinos separated by a small mass difference \cite{pDirac,Esmaili:2012ac}. In this scenario, the only new parameters
introduced are the three pseudo-Dirac neutrino mass differences, $\delta m^2_j =(\,m^+_j\,)^2-(\,m^-_j\,)^2$. While such neutrinos are
indistinguishable from Dirac neutrinos in most cases due to the smallness of $\delta m^2_j$, they lead to an oscillatory and
flavor-dependent reduction in flux. Flavor compositions are modified from the standard value of $\phi_\beta$ by
the amount $\delta \phi_\beta = - \Delta_\beta \,\phi_\beta$ with
\bea
\Delta_\beta = |U_{\beta 1}|^2 \;\chi_1+|U_{\beta 2}|^2 \;\chi_2+|U_{\beta 3}|^2 \;\chi_3\,,
\eea
where $\chi_j=\sin^2(\,\delta m^2_j \,L/4\,E)$ can be either $\frac12$ or $0$ after statistical average, depending on
whether $\delta m^2_j$ is accessible or not. The track-to-shower ratio becomes
\bea
R' = \frac{p_{{\rm CC}}\,(1- \Delta_\mu)\,\phi_\mu }{p_{{\rm NC}}\,(1- \Delta_\mu)\,\phi_\mu+(1- \Delta_e)\,\phi_e + (1- \Delta_\tau)\,\phi_\tau}\,.
\eea

\begin{figure}[t!]
\includegraphics[height=6.5cm, width=7.5cm]{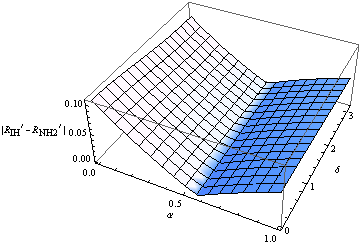}
~~
\includegraphics[height=6.5cm, width=7.5cm]{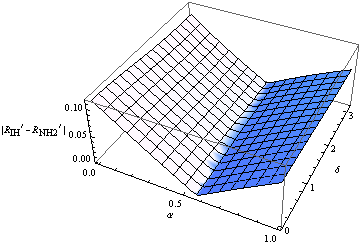}
\caption{The difference $|R'_{\textrm{IH}} - R'_{\textrm{NH2}}|$ as a function of both $\alpha$ and $\delta$ for pseudo-Dirac neutrinos.
The left and right plots correspond to the parameter sets $\{\,\chi_1=0,\, \chi_2=\frac12,\,\chi_3=\frac12\,\}$ and
$\{\,\chi_1=\frac12, \,\chi_2=\frac12,\,\chi_3=0\,\}$ respectively.}
\label{fig:3Dpd}
\end{figure}

We have explored different combinations of $\{\,\chi_1,\, \chi_2,\,\chi_3\,\}$ by studying
3D plots with axes being $\alpha$, $\delta$ and $R'$.
For $\{\,\chi_1=\frac12$, \,$\chi_2=0 ~\, \textrm{or} ~\, \frac12, \, \chi_3=0\,\}$ and
$\{\,\chi_1=0$, \,$\chi_2=0 ~\, \textrm{or} ~\, \frac12, \, \chi_3=\frac12\,\}$, we find
$R'> R$ and $R' < R$ respectively. The range of enhancement and reduction with respect to $R$ can be summarized as
$0 \lesssim |\, R'- R\,| \lesssim 0.1$. For $\chi_1 = \chi_3$, we obtain $R' \approx R$.
The above statements are valid for any injection model, mass hierarchy and $\delta$.

In Fig. \ref{fig:3Dpd}, we display the 3D plots corresponding to
$\{\,\chi_1=0,\, \chi_2=\frac12,\,\chi_3=\frac12\,\}$ and
$\{\,\chi_1=\frac12, \,\chi_2=\frac12,\,\chi_3=0\,\}$. Comparing these two plots with Fig. \ref{fig:3Ddiff}, one can see that
$|R'_{\textrm{IH}} - R'_{\textrm{NH2}}| > |R_{\textrm{IH}} - R_{\textrm{NH2}}|$ and the difference could be as large as 0.03.
Thus, for these two cases, pseudo-Dirac neutrinos require lower sensitivities at neutrino telescopes to distinguish NH2 from IH.
For all other combinations of $\{\,\chi_1,\, \chi_2,\,\chi_3\,\}$, we find that $|R'_{\textrm{IH}} - R'_{\textrm{NH2}}|$
has almost the same magnitude as $|R_{\textrm{IH}} - R_{\textrm{NH2}}|$ for any given injection model and $\delta$.
In other words, the corresponding 3D plots for $|R'_{\textrm{IH}} - R'_{\textrm{NH2}}|$ in these cases appear very similar to
the one shown in Fig. \ref{fig:3Ddiff} and so we do not display them.


\section{Conclusions}

In this paper, we have studied the cosmic neutrino flavor ratios against the undetermined Dirac CP-violating phase at neutrino telescopes.
As a consequence, we have demonstrated how to probe mass hierarchy at neutrino telescopes by the precise measurements of the
cosmic neutrino flavor ratios. Our scheme is based on the most updated global neutrino data fitting by \cite{Valle:2014} whose data suggest the possibilities of NH1 with $\theta_{23} > \pi/2$, NH2 with $\theta_{23} < \pi/2$
and IH with $\theta_{23} > \pi/2$.

We have investigated the pion-chain, damped-muon and beta-beam injection models in detail.
Since it is possible that there are some deviations from these three idealized models, we have also
studied the general injection model parameterized by Eq. \eqref{general}. We have shown that unless the injection model
has roughly equal $\nu_e$ and $\nu_\mu$ initial compositions ($\alpha \approx 1/2$), we should be able to distinguish
NH2 from IH when the neutrino telescopes could measure the track-to-shower ratio $R$ with the sensitivities down to about 0.02 or lower.
The sensitivities required are independent of the undetermined values of the Dirac CP-violating phase.

Moreover, we have explored the possible effects of active-sterile mixing, neutrino decay and pseudo-Dirac nature of neutrinos.
Since the active-sterile mixing is small, our scheme is completely not affected by it. A distinctive feature of neutrino decay
is that if we observe $0.05 < R < 0.14$ ($R\sim 0.65$), it would strongly indicate neutrino decay with NH (IH), regardless of the undetermined values of $\delta$. If neutrinos are pseudo-Dirac, there are many possibilities. However, for most of the possible combinations of $\{\,\chi_1,\, \chi_2,\,\chi_3\,\}$, the sensitivities at neutrino telescopes required to distinguish NH2 from IH are almost the same as those in the standard scenario for any injection model and $\delta$. The only exceptions are $\{\,\chi_1=0,\, \chi_2=\frac12,\,\chi_3=\frac12\,\}$ and $\{\,\chi_1=\frac12, \,\chi_2=\frac12,\,\chi_3=0\,\}$ where the required sensitivities are lower than those in the standard scenario.


Finally, a critical discussion about our work is necessary. In this paper, our definition for $R$ in Eq.\eqref{DefineR} is an idealized one.
For instance, it has neglected the fact that for NC events, only a fraction of the incident neutrino energy is deposited into the shower in the detector. To account for the difference between the incident and deposited neutrino energies, one would also have to take into account the power spectrum of the astrophysical neutrino flux. However, the power spectrum of the astrophysical neutrino flux has its own uncertainties.
We note that the main purpose of this paper is to provide an \emph{order-of-magnitude} estimation of the sensitivity required to probe mass hierarchy at future neutrino telescopes. It is most likely that IceCube cannot achieve this sensitivity. Nevertheless, the proposed expansion of IceCube \cite{IceCubeGen2} and the soon-to-be deployed KM3NeT \cite{KM3net} may perhaps reach this sensitivity. A more precise calculation for $R$
is apparently more trustworthy (which is beyond the scope of the current paper), but given that there are many uncertainties ahead of us (in 15
years or more), we believe that an \emph{order-of-magnitude} estimation is sufficient for our purpose.

\acknowledgments

We thank Markus Ahlers, Tyce DeYoung, Yu Gao, Vikram Rentala, Sergei Sinegovsky, Mariam T\'{o}rtola and Tom Weiler for useful conversations. L.F. and C.M.H. were supported in part by the Department of Energy (DE-FG05-85ER40226) and Office of the Vice-President for Research and Graduate Studies at Michigan State University respectively.


\end{document}